%% file: ms.tex
\documentclass[12pt,preprint]{aastex}
\input epsf
\usepackage{rotating}
\usepackage{float}
\shorttitle{Metastable Dynamo}
\shortauthors{Brown}

\begin{document}
\title{The Metastable Dynamo Model of Stellar Rotational Evolution}

\author{Timothy M. Brown}
\affil{Las Cumbres Observatory Global Telescope Network, 6740 Cortona Dr. Suite 102, Goleta, CA 93117, USA;  tbrown@lcogt.net}
\begin{abstract}
This paper introduces a new empirical model 
for the rotational evolution of Sun-like
stars --  those with surface convection zones and non-convective interior
regions.
Previous models do not match the morphology of observed 
(rotation period)-color  
diagrams, notably the existence of 
a relatively long-lived ``$C$-sequence'' of fast
rotators first identified by \citet{barnes2003}.
This failure motivates the Metastable Dynamo Model (MDM) described here.
The MDM posits that stars are born with their magnetic dynamos operating
in a mode that couples very weakly to the stellar wind,
so their (initially very short) rotation periods at first 
change little with time.
At some point,
this mode spontaneously and randomly changes to a 
strongly-coupled mode, the transition occurring with a mass-dependent 
lifetime that is
of order 100 MYr.
I show that with this assumption, one can obtain good fits to observations of 
young clusters, particularly for ages of 150 MYr to 200 MYr.
Previous models and the MDM both give qualitative agreement with the
morphology of the slower-rotating ``$I$-sequence'' stars,
but none of them have been shown to accurately reproduce the 
stellar-mass-dependent
evolution of the $I$-sequence stars, especially for clusters older than a few
hundred MYr.
I discuss observational experiments that can test aspects of the MDM,
and speculate that the physics underlying the MDM may be related
to other situations described in the literature, in which
stellar dynamos may have a multi-modal character.

\end{abstract}


\parindent=10pt

\section {Stellar Rotation:  Phenomenology and Theory}
In recent years, improved observational methods have drawn attention 
to subtle features of stars, notably the
rotational period, and the strength and character of stellar magnetic
activity.
Observational progress has been particularly
marked in relation to stellar rotation, and to the asymmetric surface structures
(magnetic in origin) that are its visible symptoms.
Rotational periods can be measured accurately by observing the brightness
modulation as spotted stars spin.

\begin{figure}[!hb]
    \epsscale{0.75}
    \includegraphics[angle=90,scale=0.7]{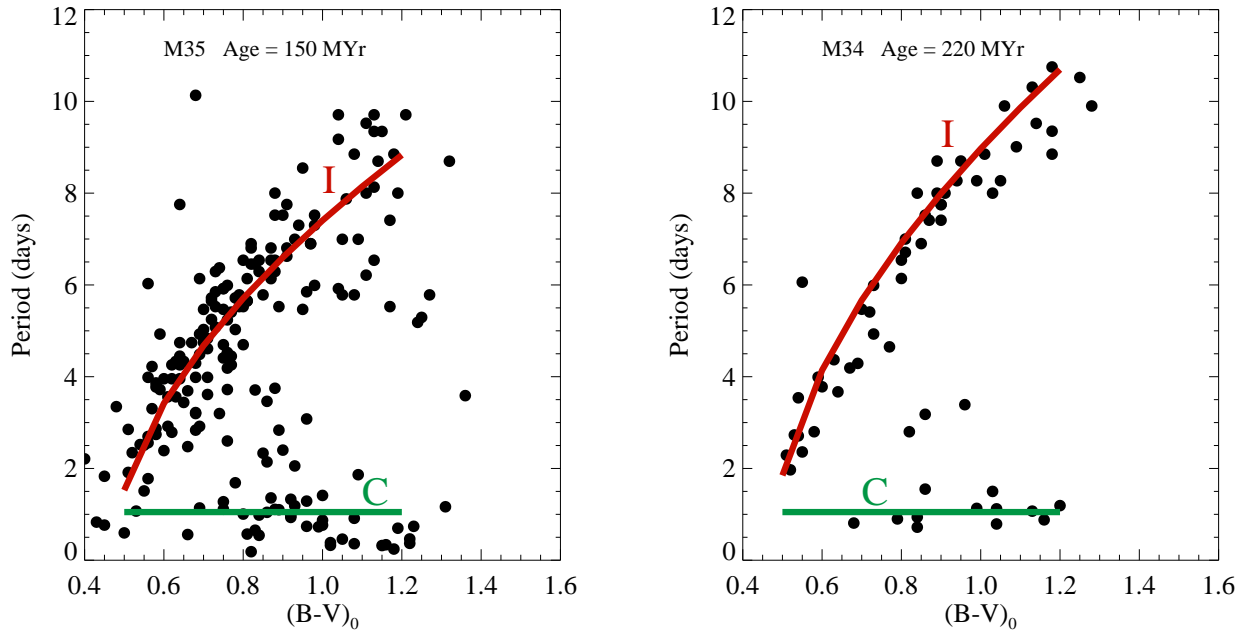}
    \figcaption[fig1.eps]{Rotational period $P_{rot}$ 
plotted against dereddened $(B-V)$
color, for the intermediate-age clusters M35 (left) and M34 (right).
Most stars lie on one of two principal sequences, identified as $C$ and $I$
by \citet{barnes2003}, although some stars are found in the intervening gap.
Data are from \citet{meibom2009, meibom2011a}.}
    \protect\label{Figure 1}
\end{figure}

Figure 1 shows such rotation periods $P_{rot}$ (days),
plotted against 
dereddened $(B-V)$ color for stars in the clusters M35 
\citep{meibom2009}, age about 150 MYr,
and M34 \citep{meibom2011b}, age about 220 MYr.
The stars plotted here (and considered in the rest of this paper) are
all Sun-like, in the sense that they have vigorous surface convection
zones (CZs), with radiative zones underneath.
Thus, these stars span spectral types from mid-F to early-M.
The current paper is not meant to apply to the late M-type
stars, because these are thought to be fully convective, and may well
be governed by different physics than for hotter stars.
The $P_{rot}$ data plotted in the Figure come from period analysis of the
brightness modulation of rotating spotted stars.
The stars with measurements are therefore those that display evidence of
asymmetrically distributed starspots.
Moreover, 
the stars represented in the Figure have been carefully vetted,
using radial velocity data and other criteria, to reject
non-members of the respective clusters.
The question naturally arises whether these visibly-spotted stars, 
which typically comprise about half of the cluster stars observed, are
representative of the cluster population.
Fortunately, spectroscopic observations of cluster stars with and without
starspot modulation indicate that there are no systematic differences
between these two groups \citep{jackson2012}.

Of primary interest in the Figure are two relatively tight sequences
of stars (labeled $I$ and $C$, after \citet{barnes2003}).
The $I$-sequence consists of relatively slow rotators.
With increasing age $t$, the $I$-sequence becomes more tightly defined,
and moves to longer periods (roughly $P_{rot} \propto t^{1/2}$), as the outflow
of magnetized stellar winds drain angular momentum from the stars.
Stars on the $C$-sequence are rapid rotators.
In young clusters $t \leq 150$ MYr), the $C$-sequence has members that 
span the mass range of Sun-like stars.
In older clusters the $C$-sequence erodes away, starting with the higher-mass
stars.
By the age of the Hyades (600 MYr), the only remaining fast rotators
are M-type dwarfs.  
Between the sequences (the ``gap''), the $P_{rot}$-color diagram shows
a paucity of stars, but it is by no means empty.

Since the 1980s, considerable theoretical work has gone into modeling
the processes leading to the behavior illustrated in Fig. 1.
In the next section I will outline the conceptual basis
for three such models, one of which, the Metastable Dynamo Model (MDM),
is  described here for the first time.
All of these incorporate magnetic wind braking to slow stellar rotation
with age, but they differ in respect to other processes that may be
acting within stars.
Anticipating later results, I find that all three models reproduce
the mass- and time-dependence of the $I$-sequence in general terms,
but that all have difficulty doing so in detail, over the range of ages between
tens of MYr and about 1 GYr.
It seems plausible, however, that more elaborate future models of 
each of the three sorts may properly capture the behavior of the $I$-sequence.
The main improvement offered by the MDM lies in its description of the
morphology and time dependence of the $C$-sequence and of the gap at longer
rotation periods.
Indeed, improving this description was the original motivation for the MDM,
and is the principal focus of the rest of this paper.

\section{Current Models of Rotational Evolution}

\subsection{Double Zone Model}
The prevailing model of the evolution of stellar rotation is what I term
(following \citet{spada2011})
the Double Zone Model (DZM).
It has evolved over the last 25 years from the stellar wind torque law
in the form written by \citet{kawaler1988}, with still earlier progenitors 
including \citet{weber1967, mestel1968} and \citet{belcher1976}.
Since the early 1990s, the model has been extended and tested against
successively improved observations in a series of papers,
e.g. \citet{pinsonneault1990, macgregor1991, krishnamurthi1997, irwin2009, 
denissenkov2010, epstein2014}.
It has recently been generalized in several ways, e.g. by \citet{reiners2012, spada2011} and \citet{gallet2013}.
Stripped of most physical justification, the model may be
summarized as follows:

(1) The torque acting on a star because of its magnetized stellar wind
is given by the bifurcated expression 
\begin{eqnarray}
{dJ \over dt} \ = \ K_W \ \Omega^3 \ \left [ {R \over M} \right ]^{1/2},
\ \ \  \ \Omega \leq \Omega_{crit}, \\
\ \ \ \ \ \ \ \ \  = \ K_W \ \Omega \Omega_{crit}^2 \left [ {R \over M} \right ]^{1/2},
\ \ \ \Omega \geq \Omega_{crit}, \ \ \ \nonumber
\end{eqnarray}
where $J$ is the star's total angular momentum, $M$ and $R$ are the star's
mass and radius (both in solar units), and $\Omega$ is the stellar rotation
frequency, often expressed in units of the Sun's rotation frequency
of about $3 \times 10^{-6}$ rad s$^{-1}$.
The constant $K_W$ has the dimensions of a moment of inertia (g cm$^2$)
and is chosen to give the solar rotation frequency
at the solar age.
This expression (with torque scaling as $\Omega^3$ for slow rotators)
yields the \citet{skumanich1972} $\Omega \propto t^{-1/2}$ rotation law for old stars.

(2) Initial conditions are applied at an age of 1 to 20 MYr
counting from the birth line of \citet{palla1990}, 
corresponding to current estimates of the time during which contracting
protostars are magnetically locked to their natal disks.
Initial rotation periods $P_0$ are usually in the range 1 to 15 days,
corresponding to the period distributions seen in the youngest
open clusters (e.g., \citet{rebull2001, rebull2004}).

(3) The dynamo saturation frequency $\Omega_{crit}$ is actually 
taken to be a function of stellar mass $M_*$
(or equivalently of $(B-V)$ color).
Its effect in the model is to decrease the torque acting on fast-rotating 
(periods of a few days or less) stars, so their short rotation periods can
survive long enough for us to see them 
in clusters with ages as great as 500 MYr.

(4) The angular momentum in stellar convection zones is assumed to be 
weakly coupled to that in their radiative interiors.
This allows the CZ to rotate more or less independently of the star's
interior.
The coupling is characterized by an equilibration timescale $\tau_c$,
which is a function of stellar mass, increasing from 10 MYr for stars
with more than solar mass, to greater than 100 MYr for early M-type stars.
In addition, to avoid spoiling the fit to fast-rotating stars, $\tau_c$
must depend also on $\Omega$ itself, being 1 MYr or less for rapid rotators,
and attaining the values just mentioned only for slow rotators
\citep{denissenkov2010}.
Thus $\tau_c$ is, in principle, a free function of mass and $\Omega$. 
As a simplification, \citet{denissenkov2010} took $\tau_C$ to be short
(1 MYr) for stars rotating faster than a critical period $P_{crit}$;
otherwise $\tau_C$ was taken to be long, of order 100 MYr.
In either case,
the equilibrium rotation frequency of the CZ is determined by balancing
the angular momentum flux across the bottom of the CZ with that lost
to the stellar wind.
  
Recently \citet{reiners2012} and \citet{gallet2013} have proposed variants
on the DZM, differing mainly in the torque laws assumed.
\citet{reiners2012} argue that the relevant magnetic field quantity is the
star's typical surface magnetic field strength, not its magnetic flux
(as assumed by, e.g., \citet{kawaler1988}).
This identification results in a strong $R^{16/3}$ dependence of the torque on 
the stellar radius, and leads to a time dependence of $P_{rot}$ that
scales as $\Omega$ for $\Omega$ faster than some critical value, and
as $\Omega^5$ if $\Omega$ is slower than that value.
For a solar-mass star, this torque law generates an evolution of $P_{rot}$ that
resembles the Skumanich $t^{1/2}$ law, but only in the sense of the average
behavior over several GYr.
For other masses and ages up to a few GYr, a Skumanich-like power law is
not obtained.
\citet{gallet2013} derive a torque law from numerical stellar wind simulations
by \citet{matt2012}.
Their torque law has no simple analytic form, but it does lead to a 
fairly strong dependence of the torque on $R$.
In the limit of slowly-rotating stars its dependence on $\Omega$ is 
slightly stronger than $\Omega^3$,
while the rapid-rotation limiting dependence is slightly weaker than $\Omega$.

\subsection{Barnes's Symmetrical Empirical Model}
\citet{barnes2010} and \citet{barneskim2010} sought to describe observed
cluster $P_{rot}$-color diagrams empirically,
connecting their expressions with astrophysical concepts only after
obtaining a satisfactory fit to the observations.
For purposes of this paper, I call their model the Symmetrical Empirical
Model (SEM).
Its logic runs as follows:

(1) Stars on the $I$- and $C$-sequences obey different period-evolution equations,
namely
\begin{eqnarray}
P_I(B-V, t) \ = \ f(B-V) g(t), \ \ (I-{\rm sequence}) \\
P_C(B-V, t) \ = \ P_0 e^{[t/T(B-V)]}. \ \ (C-{\rm sequence}) \nonumber
\end{eqnarray}
Here $g(t) \ = \ t^{-1/2}$, as in the Skumanich law, $P_0$ is a constant
period, and the functions $f(B-V)$ and $T(B-V)$ are initially arbitrary
functions that can be determined from fits to the observations.
Both $T(B-V)$ and $f^2(B-V)$ have dimensions of time.
\citet{barneskim2010} note that both these functions appear to
be related in simple ways to a physically interesting quantity,
namely the turnover time of the convection zone, denoted $\tau$.

(2) Accepting this identification, 
the period evolution expressions can
be combined into one period evolution equation that has the correct
behavior both for large and small values of the period $P$:
\begin{equation}
{dP \over dt} \ = \ \left [ {k_I P \over \tau} \ + \ {\tau \over k_C P}
\right ]^{-1} \ ,
\end{equation}
where $k_I$ and $k_C$ are dimensionless constants 
determined from fits to the data.
This equation gives the SEM its name.
To parallel the development of the DZM, I transform the period evolution
equation to a torque law, assuming solid-body rotation, so that the relevant
moment of inertias for both sequences are equal to the 
total stellar value $I_*$.
The resulting torque law is
\begin{equation}
{dJ \over dt} \ = \ - I_* {\Omega^2 \over 2 \pi } \left [ {{2 \pi k_I} \over
 {\tau \Omega}} \ + \ {{\tau \Omega} \over {2 \pi k_C}} \right ] ^{-1}.
\end{equation}

(3) Initial conditions may be applied as with the DZM, though \citet{barnes2010}
prefers to set initial periods at the zero-age main sequence (ZAMS), roughly
50 MYr after the birth line.
At this time the rotation periods are at their shortest, limited at the
short-period end by the breakup equatorial speed of the star.

The SEM is a descriptive, not explanatory model, as \citet{barneskim2010}
take pains to point out.
That is, the physical processes that determine the functions $f(B-V)$
and $T(B-V)$ are not specified.
The possible connection between these functions and the convective
turnover time $\tau$ is, however, suggestive.
If the reality of this relationship can be verified, then it might place a
useful constraint on more physics-based models of stellar magnetic fields.

\section{Computing the Rotation Evolution with Time}

Based on the prescriptions in the previous section, it is fairly
straightforward to compute the time evolution of $P_{rot}$ at the surface
of a star, according to either the DZM or the SEM.
In addition to the torque equations given above, for this purpose 
one requires the time
history of the radii and moments of inertia of the stellar radiative
zone and outer convective envelope.
I computed these quantities for a set of 11 models with initial masses
of \{0.3, 0.4,....1.3\} $\times \ M_{\sun}$, using the MESA suite of
stellar evolution codes \citep{paxton2011}.
These were standard (non-rotating) models using solar abundances and
default settings for all microphysics processes.
As discussed by, $e.g.,$ \citet{denissenkov2010}, gross structural
differences between
rotating and non-rotating stars are small except for $\Omega$ near the
break-up value, and are not of qualitative importance for the rotational
evolution.
For masses intermediate between the computed models, I obtained the needed
radii and moments by linear interpolation between models.
Because the MESA timesteps were not always small enough for stable numerical
integrations of the rotation, I also interpolated the stellar data onto a finer
time grid, typically involving 3000 time steps to span the Sun's age,
roughly equally spaced in $\log t$. 
For a direct comparison of my computed results with published ones,
I computed $(B-V)$ and the convective turnover time $\tau$
from the tabular functions of mass (evaluated at an age of 500 MYr)
given by \citet{barneskim2010}).
($\tau$ is plotted in their Fig. 9.)

Lastly, one requires initial conditions, namely the interval $\tau_{disk}$
between the
\citet{palla1990} birth line and the time when the star unlocks 
from the protostellar disk,
and $P_0$, the star's rotation period at unlocking.
In the following simulations I set $\tau_{disk}$ to either 5 or 10 MYr,
depending on which published results I sought to emulate.
In all cases, I drew $P_0$ values from a distribution intended
to mimic the observed rotation periods 
(independent of stellar mass) in the Orion Nebula Cluster (ONC),
as compiled in the Open Cluster Database \citep{mermilliod1995},
drawing data from \citet{walker1990, eaton1995, choi1996} and 
\citet{rebull2001}.
Fig. 2 shows this initial period distribution. 
It develops, however, that there is no qualitative difference between the
population synthesis results for the ONC $vs.$ a uniform distribution
between $P_0$ of 1 to 15 days,
for cluster ages above about 100 MYr.
As elaborated below, this is a result of the convergent nature of the
$\Omega^3$ torque law ascribed by all three model classes (DZM, SEM, MDM)
to stars of greater age.
For population synthesis computations, I usually drew stellar masses
from a uniform distribution spanning the range of masses in the MESA
models.
When attempting to match model parameters to specific cluster observations,
however, I drew masses from distributions constructed to reproduce
the observed number of stars $vs.$ $(B-V)$ color.

\begin{figure}[!hb]
\begin{center}
\includegraphics[angle=90,scale=0.5]{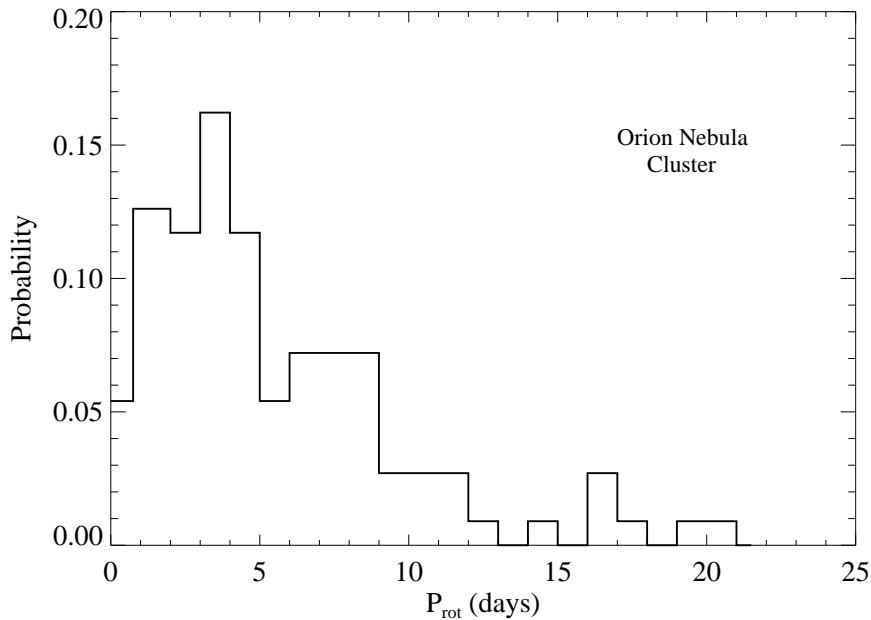}
\figcaption[fig2.eps]{Initial period $P_0$ probability distribution assumed
for all population synthesis models in this paper.
$P_0$ is the rotation period that applies at the ``birth line'' of
\citet{palla1990},
and also (by definition) the period at the end of the disk-locking time
$\tau_{disk}$ for the particular star or model in question.
This distribution is derived from the Open Cluster Database
\citep{mermilliod1995}, with original data from
\citet{walker1990, eaton1995, choi1996, rebull2001}.
\label{Figure 2}}
\end{center}
\end{figure}

\section{Comparing the DZM and SEM to $P_{rot}$-color Diagrams}

\begin{figure}[!hb]
\begin{center}
\includegraphics[angle=90,scale=.65]{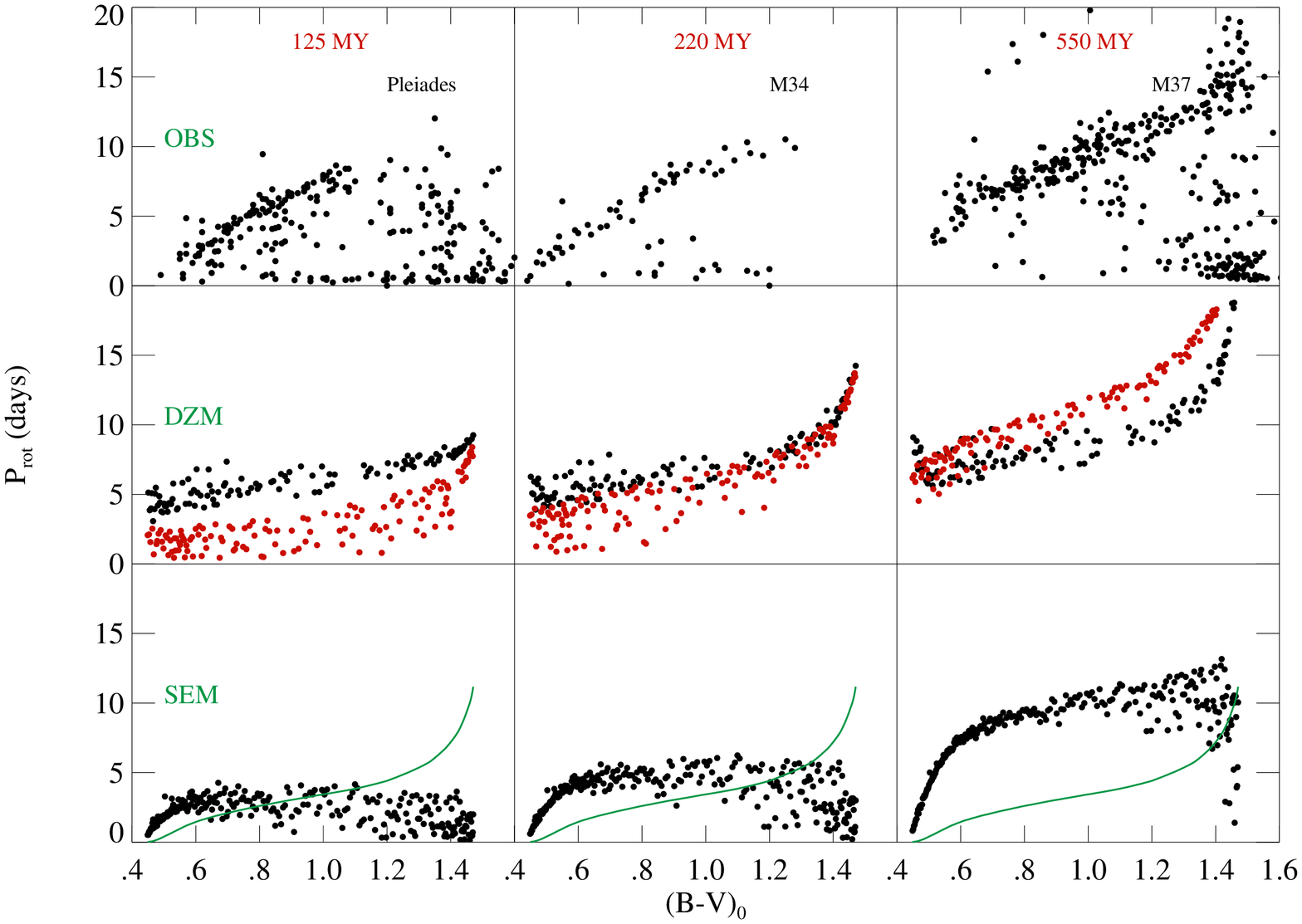}
\figcaption[fig3.eps]{Comparison of observed (top row of panels) 
$P_{rot}$-color diagram for the open clusters \{Pleiades, M34, M37\}
with the results of population synthesis for 
the DZM (middle row), and SEM (bottom row) models.
Observations are from 
(Pleiades: \citet{hartman2010}),
(M34: \citet{meibom2011a}),
(M37: \citet{hartman2009}).
Both models are calibrated to give the solar rotation rate for a 1-$M_{\sun}$
star at the solar age, and all start with 250 stars 
having $\tau_{disk}$ = 5 MYr, 
initial periods distributed as in the Orion Nebula Cluster,
and masses uniformly distributed over $0.4 M_{\sun} \leq
M \leq 1.2 M_{\sun}$.
The DZM parameters are in the ranges given by
\citet{denissenkov2010};
red symbols indicate stars with initial periods $P_0$ so small that they
rotate essentially as solid bodies, while black symbols
denote stars with larger $P_0$, so that their convection zones and
radiative interiors may rotate at widely different rates.
The SEM parameters agree with \citet{barneskim2010};
green curves show $P_{gap}(B-V)$, where period evolution is fastest.
Detailed parameter values are given in the text.
\label{Figure 3}}
\end{center}
\end{figure}

Fig. 3 shows the observed $P_{rot}$-color distribution for three open
clusters of different ages (top row of panels),
along with population synthesis models at the same nominal ages
using the DZM (middle row) and SEM (bottom row).
Both the DZM and SEM models are calibrated to give the solar rotation
period at the solar age, for a 1$M_{\sun}$ star.

\subsection{The DZM}

For the DZM model results, I used parameter values within the bounds suggested
by \citet{denissenkov2010}: $\tau_{disk}$ = 5 MYr,
$\Omega_{crit} = 8\Omega_{\sun}$,
$P_{0 {crit}}$ = 4 days, and
$\tau_c$ = 50 MYr for 1 $ M_{\sun}$, increasing to 200 MYr for 
0.5 $ M_{\sun}$.
\citet {denissenkov2010} did not offer a specific recipe for the functional
form of $\tau_c ({\rm mass})$;
for definiteness, I made it proportional to the convective turnover
time $\tau$.
For the ONC initial period distribution and $P_{0 {crit}}$ = 4 days,
about half of the stars rotate as solid bodies, and half display differential
rotation, with their convection zones decoupled from their interiors.
In the Figure, the solid body rotators (henceforth SB) are shown in red, 
and the differential rotators (henceforth DR) in black.

With suitable parameter choices,
the DZM reproduces the top and bottom
10th percentile points of the surface rotation distributions in 16
open clusters of various ages \citep{denissenkov2010}.
Other studies, e.g. \citet{irwin2007, irwin2011, gallet2013} obtain
similar agreement with various data sets.
Note, however, that to get this agreement, both $\tau_{disk}$ 
and $P_0$ must be 
large to match the slow rotators, and both must be small for the fast rotators.
Moreover, the studies just mentioned judge the quality of their fits
by comparison with specified percentile points in the distribution
of $P_{rot}$ (often summed over stellar mass) 
at each value of the estimated stellar age.
This parameterization of the observed distributions in terms of
percentiles is of course a good
starting point, but it loses much of the information present in the
observations, and can even be misleading when applied to bimodal distributions
such as those seen in young clusters. 
To capture all the information present in current $P_{rot}$-color diagrams,
it is better to compare the full diagrams (at several ages) to
population synthesis models, or to compare observed and model probability 
distributions (with resolution in both mass and age) derived from such
diagrams.

In the current population synthesis with $P_0$ distributed in accord with
observations of the ONC, the agreement of standard DZM models
with observed $P_{rot}$-color
diagrams is not good.
At 125 MYr, the SB and DR groups form two
distinct sequences, with the SBs spinning faster.
But the SBs move to longer periods faster than the DRs,
so that by 220 MYr the gap between the seqences disappears,
and the two sequences merge into one.
Also by this age the true C-sequence stars (those with periods shorter
than about 2 days) are found only at larger masses;  none appear
for masses below about 0.9 $M_{\sun}$.
This contradicts the observed behavior that rapid rotators disappear first
at larger masses.
At later ages $dP_{rot}/dt$ for the SBs continues to be larger than for
the DRs, so that by 550 MYr there are
once again two sequences,
but at this age the SBs have longer periods than the DRs, and neither
group contains any stars with periods shorter than about 5 days.

\begin{figure}[!hb]
\begin{center}
\includegraphics[angle=90,scale=0.65]{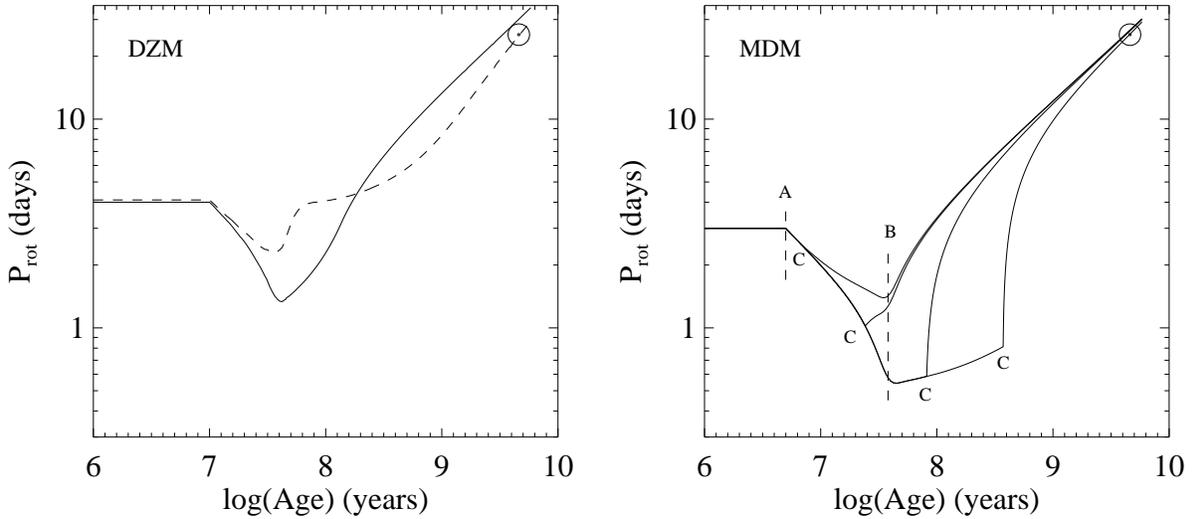}
\figcaption[fig4.eps]{Left:  Evolution of two models according to
the DZM.
Both models are for 1 $M_{\sun}$ and identical initial 
rotation conditions $P_0$
and $\tau_{disk}$.  The solid curve shows solid body
(SB) rotation, while the dashed curve shows the envelope $P_{rot}$
for a model with radial differential (DR) rotation.
The Sun symbol shows the solar age and $P_{rot}$ values. 
Right: Rotational evolution for four MDM models.
All correspond to 1 $M_{\sun}$ and to $P_0 = 3$ days, $\tau_{disk} = 5$ MYr.
The vertical dashed line {\bf A} corresponds to disk unlocking,
{\bf B} to the star's arrival on the main sequence,
and the lines {\bf C} to transitions from the weakly- to strongly-coupled
dynamo modes occurring at \{10, 30, 100, and 300\} MYr (left to right).
Note that in the strongly-coupled mode, angular momentum loss may be significant
during the pre-main-sequence phase.
\label{Figure 4}}
\end{center}
\end{figure}

The crossing behavior of the SB and DR sequences results from the smaller
torque hypothesized to act on the members of the DR sequence,
this torque being determined by angular momentum transport between the
stellar interior and the envelope,
rather than between the envelope and the wind.
By disconnecting the interior from the wind torque, the differential
rotation hypothesis allows models in which the surface rotation rate
first decelerates quickly, 
and then stalls at a near-constant value for an extended
time (e.g. \citet{krishnamurthi1997, denissenkov2010}).
Meanwhile, the torque acting on the SB stars is larger than on the DRs
of similar mass,
typically by large factors.
Thus, eventually the SBs spin down to periods that are longer than
those of the DRs.
For equal mass stars, this generates crossing paths 
in $P_{rot}$-age space, as shown in
Fig. 4.
The only evident way to avoid such crossings (and hence to preserve an
age-independent $I$/$C$ sequence morphology) is to consider groups of stars 
for which the SB/DR distinction necessarily entails large 
differences
in $P_0$ and/or $\tau_{disk}$.
(Fig. 8 of \citet{denissenkov2010} shows such an example.)
But in a population synthesis involving continuous distributions of $P_0$
and $\tau_{disk}$, such correlated behavior among the various initial values
would require new rules in a model prescription that is already 
rather complicated.
Alternatively, better agreement between observations and theory can likely
be obtained by suitably tuning the distribution of $P_0$, 
by modifying the stellar wind torque law \citep{reiners2012, gallet2013},
or through a different prescription for determining the strength of the
core/envelope coupling \citep{spada2011}.
Thus, it seems possible that better fits to the $I$-sequence
behavior could found by hypothesizing continuous transitions between
limiting values of, e.g., $P_0$ or the core-envelope coupling time $\tau_c$,
rather than the bimodal behavior that underlies my simulations shown in
Fig. 3.
So far, however, little theoretical work has been done on the distributions
needed for these parameters, or on how these might scale with stellar mass,
age, or other relevant drivers.
The range of such possibilities is therefore wide, and
moreover variation of these parameters is not obviously relevant 
to my primary concern -- the behavior of the $C$-sequence
rapid rotators.
Investigating whether modifications of the DZM might provide an
acceptable representation of the data is therefore an interesting topic,
but one that is beyond the scope 
of this paper.
I note only that DZM parameters suggested in the literature to date
seem to lead to an unsatisfactory comparison with observations,
motivating a search for alternatives.

\subsection {The SEM}

By construction, the Symmetrical Empirical Model  \citep{barnes2010, barneskim2010}
attempts to match the high- and low-$P_{rot}$ envelopes of 
observed clusters.
But (given a set of initial periods) it also predicts the period distribution 
of stars in the gap between the $C$- and $I$-sequences.
As noted by \citet{barnes2010}, 
for each mass there is a period $P_{gap} \ = \ \tau/{\sqrt{
k_I k_C}}$, for which $dP/dt$ reaches a maximum.
Using the computed mass dependence of $\tau$ and $(B-V)$, one can construct the
curve in $P_{rot}$-color space along which this maximum occurs.
This is shown as the green curve in the bottom row of panels of Fig. 3.
Via a conventional evolution/population argument, one expects the density
of stars along this curve to be smaller than at neighboring periods.
\citet{barnes2010} suggests this as an observational test of the theory,
and displays data from three clusters (the Pleiades, M35, and M37) that
provide some support for the idea.
Nonetheless, population synthesis carried out with the SEM, 
starting with the ONC
period distribution, does not produce
a clear $C$-sequence (Fig. 3, bottom row).
Although $dP_{rot}/dt$ is a maximum along the $P_{gap}$ curve,
the variation of $dP_{rot}/dt$ with $P_{rot}$ over the relevant range
of periods is not very large.
As a result the low density of stars along the $P_{gap}$ curve is difficult
to discern.
Moreover the $P_{gap}$ locus makes a clean boundary between the $C$- and
$I$-sequences only for a limited range of cluster ages.
It thus seems doubtful that the SEM's period evolution structure is responsible
for the existence of the $C$-sequence.
The SEM $I$-sequence is well defined, especially at later ages,
but overall the synthesized morphology
in the $P_{rot}$-color diagram is not a good match to observations. 

It is worth noting that \citet{barnes2010} applied $P_{rot}$ initial conditions
spanning the range 0.12 $\leq \ P_0 \ \leq$ 3.4 days 
at $t$ = 50 MYr, roughly at the ZAMS for a solar-mass star.
In my SEM simulation shown in Fig. 3, I used instead the ONC $P_0$ distribution
at the birth line.
This means that for some 50 MYr, my simulation uses the SEM equations outside
of their nominal range of validity, while 1 $M_{\sun}$  stars 
are still contracting.
However, the ONC $P_0$ distribution
evolved to an age of 50 MYr yields 0.13 $\leq \ P_{rot} \ \leq$ 3.7 days,
almost identical to the initial periods chosen by \citet{barnes2010}.
So although the pre-main-sequence phase of evolution may not be properly
represented by the SEM, the post-main-sequence behavior in my simulation
should closely mimic that of \citet{barnes2010}.

Once all the stars in a cluster have unlocked from their
disks, the SEM preserves the ordering of stars in period for each mass.
Thus, the lower envelope of the period distribution is determined
by the minimum $P_0$ found at each mass,
and $dP_{rot}/dt$ scales inversely with $\tau (M_*)$.
The effect is that rapidly-spinning stars evolve to periods longer than
(say) 2 days at younger ages for high-mass stars than for low-mass ones
(in qualitative agreement with observations).
But the evolution rate is such that for intermediate masses (corresponding
to $(B-V) \leq 1.2$ or so, all stars evolve to $P_{rot} \geq 2$ days within
less than 200 MYr, even if $P_0$ is so short that their peak equatorial 
velocity reaches the
centrifugal breakup speed.
At 200 MYr, this leads to a total absence of stars in the SEM at 
short periods and moderate masses.
There appears to be no way within the SEM to produce intermediate-mass stars
with ages above 200 MYr and $P_{rot} \leq$ 2 days, as are observed in M34
and M37.
Moreover, the SEM gives a distribution of $P_{rot}$ between its upper
and lower envelopes that is more or less uniform at all masses.
This does not resemble the observed distribution, in which the
well-populated $I$ and $C$-sequences are separated by a gap that is
sparsely populated, but not empty.

As with the DZM, it is likely that one could improve the fit between the SEM and
observations via appropriate adjustments of its parameters. 
But this too is outside the scope of the current discussion.
Rather, in the next section I describe a different model that achieves
the desired agreement more simply.

\section{Metastable Dynamo Model}
Here I introduce a third approach, which I call the 
Metastable Dynamo Model (MDM), and which
offers a new way to understand
the observed $P_{rot}$ data.
It derives from Barnes's SEM, but in 
constructing the model, I have re-interpreted his function $T (B-V)$.
The assumptions going into the MDM are:

(1) Stars on the main sequence rotate as solid bodies, with little or
no radial differential rotation.
Note that though the Sun rotates differentially in latitude, 
it is no exception to this rule;  appropriately averaged over latitude,
the Sun's convection zone rotates (so far as helioseismology can show) 
at virtually the same rate as its radiative interior \citep{gilman1989,
tomczyk1995}. 

(2) Initial rotation conditions are given by conventional disk-locking
processes, as for the DZM and SEM above.

(3) Stars spin down according to the torque law
\begin{equation}
dJ/dt \ = \ K_M  \Omega^3 f^2(B-V) ,
\end{equation}
where $f^2 (B-V)$ is the same as in Barnes's SEM model.
As in the SEM, this function contains all of the mass dependence of
the magnetized wind torque.

(4) The leading constant $K_M$ may take one of two values, 
a ``strong coupling'' one ($K_{M1}$) that is consistent with 
a solar-mass star having the solar rotation period at the solar age,
and a ``weak coupling'' one ($K_{M0}$) that is much smaller,
perhaps by a factor of 100 or more.
I tentatively identify these two values as manifestations of 
dynamo modes that are structured in different ways,
perhaps like Barnes's (2003) ``Interface''  and ``Convective'' dynamos.
This notion gives the MDM its name, but for now is a
conjecture.

(5) At the time that a star unlocks from its disk, it occupies the
weak-coupling mode.
It then spontaneously, randomly, and permanently flips to the 
strong-coupling mode.
The transition probability for this flip per unit time is given by
$1/\tau_M (M_*)$, which is essentially 
the reciprocal of Barnes's function $T (B-V)$.
This is to say, here I assume that $\tau_M (M_*)$ scales with the 
convective turnover time $\tau (M_*)$.
The assumption that all stars start in the weak-coupling mode is
arbitrary, and motivated largely by simplicity.

The time history of a star's $P_{rot}$ according to the MDM is not a 
deterministic function of its initial conditions;
the random transition between coupling modes introduces a stochastic
element into the evolution.
Several possible tracks in $P_{rot}$-age space are illustrated in the
right-hand panel of Fig. 4.
These all correspond to identical initial conditions, but with varying ages
for the transition between coupling modes.
Observe that even for late transitions (e.g., at age $\geq$ 300 MYr), 
the period convergence caused by the $\Omega^3$ torque law assures that
all stars of a given mass arrive at the solar age with very similar
rotational periods.

The MDM makes no use of the dynamo saturation frequency
$\Omega_{crit}$.
This is not to suggest that magnetic saturation does not occur;  there is strong
evidence for such an effect on the X-ray emission from magnetically
active stars, e.g. \citep{stauffer1994, pizzolato2003}.
Rather, within the MDM framework, almost all stars that rotate faster
than $\Omega_{crit}$ are ones that have not yet made the transition
from the weak-coupling mode.
For these stars, the difference between $\Omega^3$ and $\Omega \Omega_{crit}^2$
is negligible compared to the difference between $K_{M0}$ and $K_{M1}$.

\begin{figure}[!hb]
\begin{center}
\includegraphics[scale=0.5]{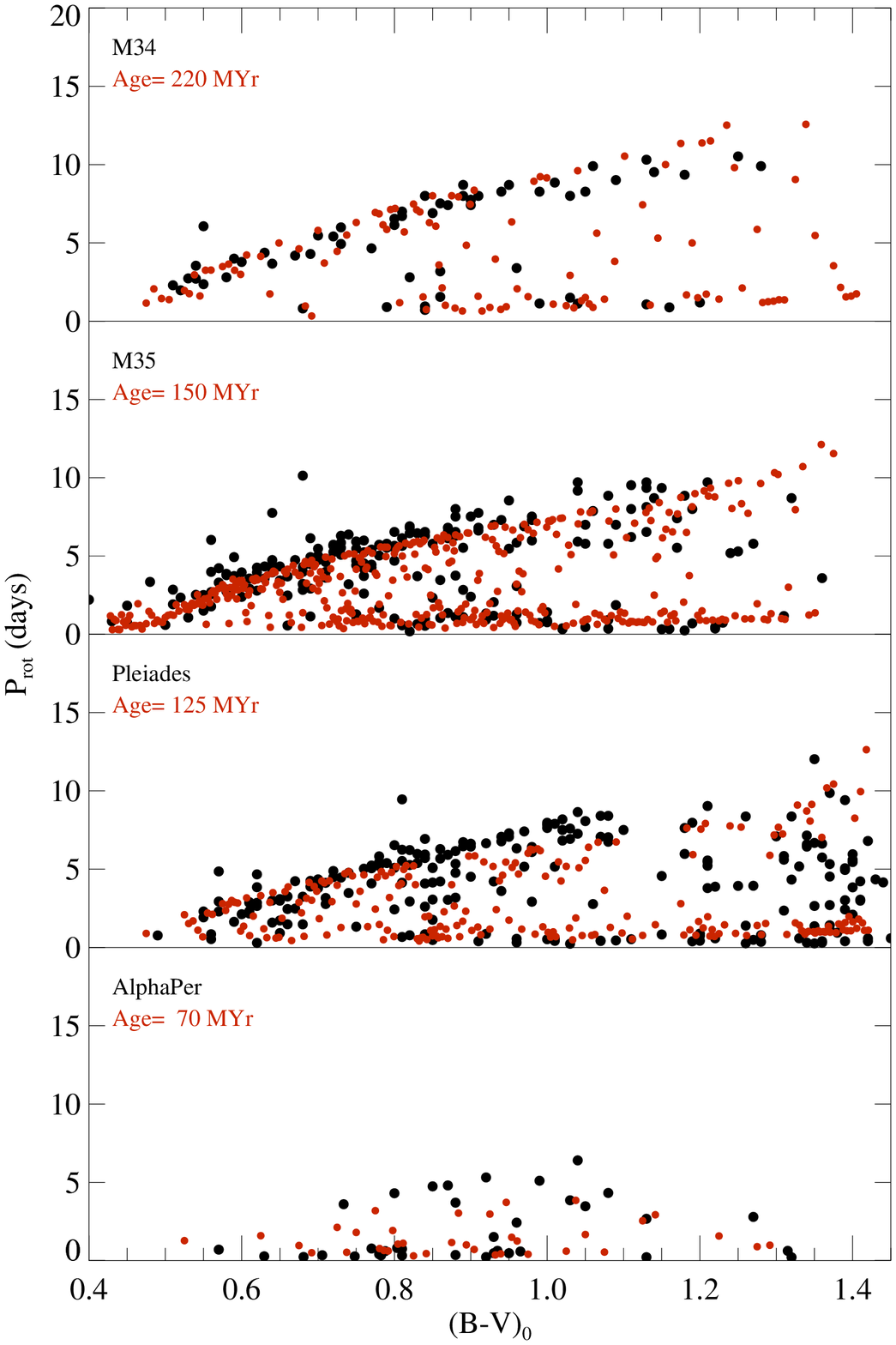}
\figcaption[fig5.eps]{Overlaid observations (black dots) and MDM
population synthesis simulations
(red dots) for clusters $\alpha$ Per, Pleiades, M35, and M34, 
(from bottom to top), approximate ages 70, 125, 150, and 220 MYr, respectively. 
Observations are from ($\alpha$ Per: as tabulated in \citet{mermilliod1995}), 
(Pleiades: \citet{hartman2010}), 
(M35: \citet{meibom2009}), 
(M34: \citet{meibom2011a}).
The MDM model parameters are those found by maximizing the joint
(over colors) Kolgomorov-Smirnov probability that model and 
observed distributions for M34 are
drawn from the same distribution ($cf.$ sections 6.1-6.2).
Parameters used for these simulations were
$\tau_{disk}=5$ MYr, $K_{M0}=5 \times 10^{29}$ g cm$^2$, 
$K_{M1}=7 \times 10^{31}$ g cm$^2$, $\tau_M =$80 MYr for stars
of 1 $M_{\sun}$, scaling as the convective turnover time $\tau$. 
The function $f(B-V)$ defining the shape of the $I$-sequence is proportional
to $\tau^{1/2}$, as in the SEM and \citet{barneskim2010}.
\label{Figure 5}}
\end{center}
\end{figure}

\begin{figure}[!hb]
\begin{center}
\includegraphics[scale=0.5]{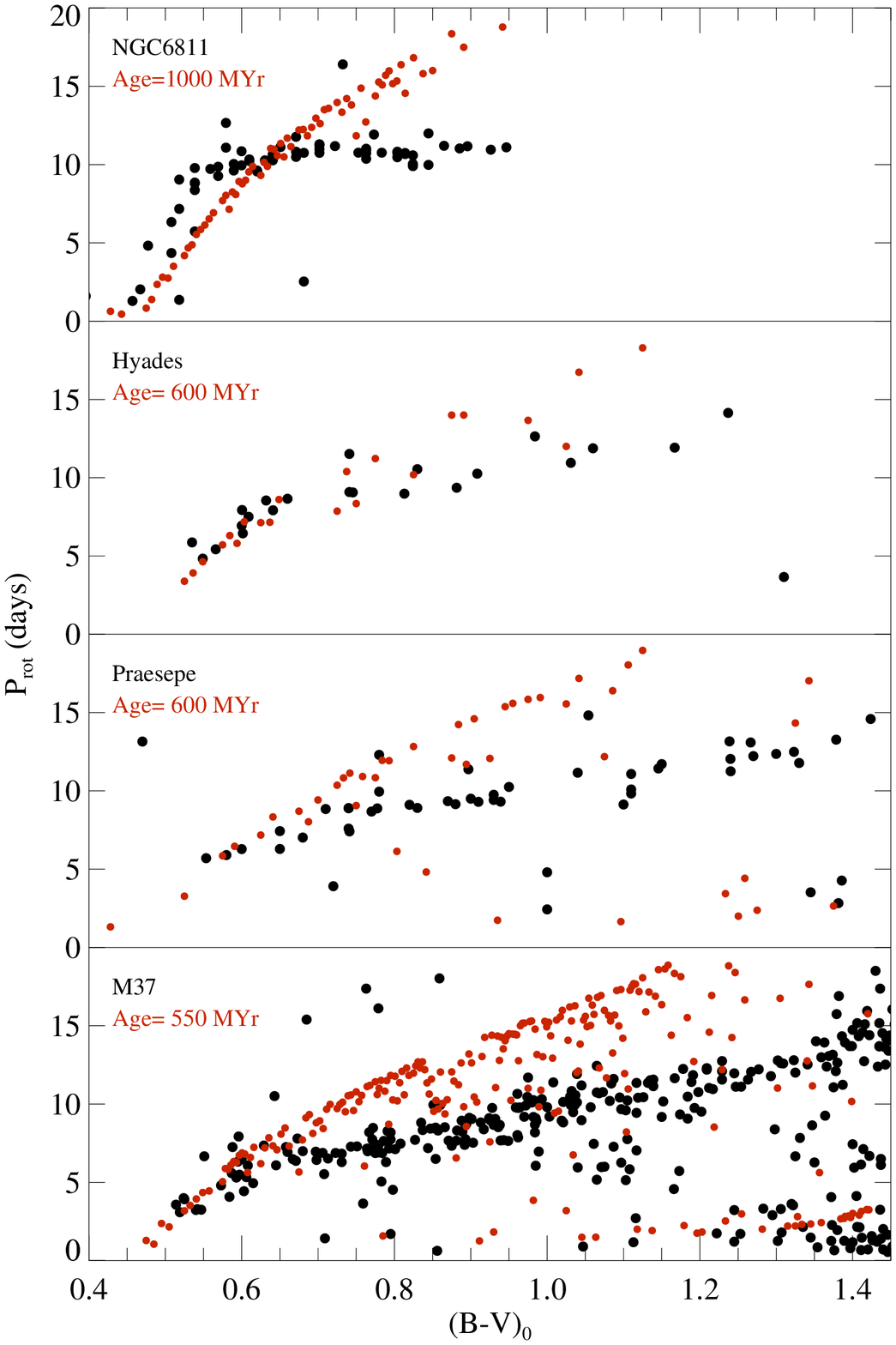}
\figcaption[fig6.eps]{Same as Fig. 5, but for the older clusters
(bottom to top) M37, Praesepe, Hyades, NGC 6811, with approximate ages
550, 600, 600, and 1000 MYr, respectively.
Observations are from (M37: \citet{hartman2009}),
(Praesepe: \citet{delorme2011}),
(Hyades: as tabulated in the Open Cluster Database \citet{mermilliod1995}),
NGC 6811: \citet{meibom2011b}). 
\label{Figure 6}}
\end{center}
\end{figure}

Figures 5 and 6 show comparisons between observations and 
MDM population synthesis models
for eight clusters of various ages.
For the simulations in these Figures, I constrained the number of model stars in
bins of similar $(B-V)$ color (width 0.1 mag) to be equal to the number
actually observed in the corresponding cluster.
Otherwise, the simulations are strictly analogous to those in Fig. 3.
Evidently the gross morphology of these simulations is much more like
the observations than for the DZM and SEM models used in Fig. 3.
The behavior of the observed $C$-sequence is captured by the models:
there are many stars in it for the younger clusters,
but by the age of the Hyades there are few or none.
Also the modeled $C$-sequence begins to thin out first at the higher masses,
as it is observed to do.
The most notable defect with respect to the $C$-sequence 
is that the models may overpopulate it at young ages;
a possibly related problem is a paucity of slow rotators at ages of
50 MYr.
Relaxing the MDM assumption that all stars are born in the weak-coupling
state would certainly help with the first of these problems, and may
also help the second.  
I have not yet tested this idea, however. 
The comparison between observed and modeled $I$-sequences
is best at around 200 MYrs;
the sequence's slope and shape are slightly incorrect for the Pleiades,
and much more so for the older clusters Hyades, Praesepe, and NGC 6811.
Similar problems occur with the DZM and SEM models, however.
All of the models act so that the shape of the $I$-sequence
remains nearly constant in time, merely scaling in magnitude with age.
The comparison shown here illustrates that this is too simple a description;
more complicated physics will have to be invoked to explain the true 
$I$-sequence behavior.

Clearly the MDM is, like the SEM, a purely descriptive model.  
In common with the DZM and SEM, it implicitly
invokes processes that have little or no quantitative physical basis
(but see recent work on bistable dynamos, {\it e.g.} \citet{morin2011, gondoin2012, gondoin2013, cook2013}.)
Nevertheless, the simulations 
show that, with plausible initial conditions, the MDM 
reproduces the appearance of observed 
$P_{rot}$-color diagrams fairly well.
At a qualitative level, this is perhaps unsurprising.
The strongly-coupled stars follow the same torque law as for the
DZM and the SEM, hence must converge to a narrow $I$-sequence
at large ages.
Similarly, the model hypothesizes weakly-coupled stars 
precisely to create a relatively long-lived $C$-sequence of short-period
rotators.
The detailed shape $P_{rot} (M_*)$ of the $I$-sequence, as well as the
number density of stars along the $C$-sequence as a function of age,
both depend on free functions of mass that are poorly motivated by theory.
Thus, the essential features of the MDM that distinguish it from the
DZM and SEM are these:
(1) Rapidly-rotating stars may for hundreds of MYr exist in a state in
which they lose very little angular momentum.
(2) When stars leave this weakly-coupled state, they quickly evolve to 
much longer $P_{rot}$.  
In other words, when it occurs, the transition in torque felt by the star 
must be large in magnitude and nearly discontinuous in time.

The qualitative success of the MDM is thought-provoking 
(though of course not conclusive in regard to physical origins).
It suggests that the idea of metastable modes in stellar magnetic evolution
may be a fruitful one.
Fortunately, there are several observational
tests that may be used to help validate or reject the model.
I will discuss these in Section 7.2 below.

\section{Quantitative Comparison of Rotation Models to Observations}

\subsection{Methodology}

To better understand the successes and failures of the rotation evolution
models, and to estimate the best parameter values for the MDM, 
it is useful to compute a quantitative measure of the goodness of fit
of the models to the cluster rotation data.
For this purpose, I used a  Kolgomorov-Smirnov (K-S) procedure.
This is similar in spirit to the approach taken by \citet{spada2011},
except that I applied it to a wider range of stellar masses,
segregated by mass (or $(B-V)$ color) into distinct subsamples.
For each model considered (DZM, SEM, or MDM with various parameter values)
and each of several $(B-V)$ color bins,
I computed the period evolution for a cohort of $10^4$ stars, uniformly
distributed in mass so as to span the desired color bin, using the
mass-color relation at 500 MYr age
as tabulated by \citet{barneskim2010},  with $P_0$ values randomly drawn
from the ONC distribution shown in Fig. 2.
To simplify the computation, I ignored the evolution of $(B-V)$ with age.
Moreover, I reduced the dimensionality
of the problem by assuming a single disk-locking age $\tau_{disk} = 5$ MYr
for all models.
These evolutions started at the \citet{palla1990} birth line, and
ran to a maximum age of 6 GYr.
For any desired age,
I could then estimate the rotational probability distribution
$\Phi_j (P_{rot}, [B-V]_j, t)$.
$\Phi_{j} $ is normalized so that 
$\int \Phi_{j}  \ dP_{rot} = 1$,
separately for each color bin $j$ and each age.
Figure 7 illustrates such probability distributions for several $(B-V)$ values,
estimated using the MDM for an age of 250 MYr. 

\begin{figure}[!hb]
\begin{center}
\includegraphics[angle=90,scale=0.6]{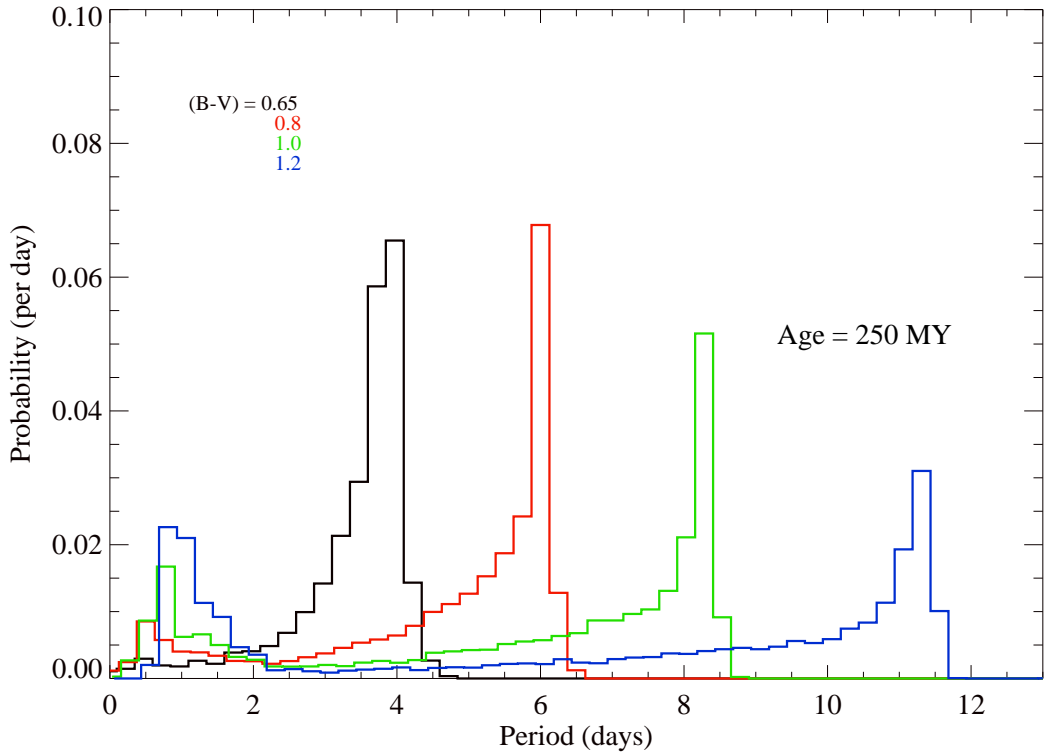}
\figcaption[fig7.eps]{MDM-derived probability distributions for
four $(B-V)$ colors, for stellar cohorts at an age of 250 MYr.
At this age, the distributions are bimodal for all $(B-V)$ 
redder than about 0.65,
with an increasingly large fraction of stars in the $C$-sequence
(roughly $P_{rot} \leq$ 2 days) as $(B-V)$ increases.
Between the short- and long-period peaks, minima in the distributions
are reached for $P_{rot}$ between 1 and 3 days, depending on color.
Stars with rotation periods near and just above these values 
have recently transitioned to the
strong-coupling mode, and are evolving to longer periods at the
maximum rate.
\label{Figure 7}}
\end{center}
\end{figure}

The model $\Phi_{j}$ distributions in Fig. 7 are useful for
illustrative purposes.
But to compare the models to observations of a particular cluster $a$, 
it is simpler
to lump each set of periods (modeled and observed) into bins 
according to $(B-V)$,
and then use the two-sided Kolmogorov-Smirnov test \citep{press1989}
to estimate for each color bin $j$ the probability $U_{aj}$ that
the two period sets are drawn from the same distribution.
To compare the success of various models in representing observations from
a given cluster $a$, I used a figure of merit defined as
$U_a \ = \ \Sigma_j \log_{10}(U_{aj})$. 
This choice corresponds to the $\log_{10}$ of the probability that
observations and simulations are drawn from the same distributions,
jointly for all of the four
color bins.
Note that since the numerical value of $U_a$ depends on the number of stars
in the $(B-V)$ bins, this figure of merit may not be used to
compare the representations of different clusters, even using the same model.
For the results described below, I used four $(B-V)$ bins, with
blue and red boundaries given by $\{(0.57,0.66), (0.67,0.81), (0.81,1.01)
(1.01,1.21)\}$.
From the 500 MYr mass-to-color relationship from \citet{barneskim2010},
this corresponds to masses of approximately
$\{(1.1,1.0), (0.99,0.89), (0.89,0.79), (0.79,0.69)\} \ M_{\sun}$.

\subsection{Optimum Parameters for MDM}

Using the  K-S measure of goodness-of-fit to the cluster rotation data,
one can place limits on the parameters that appear in the MDM
torque law.
If the physical basis of the MDM is taken seriously, then such limits
should constrain the processes whereby angular momentum is lost to the
stellar wind, as well as the nature of the transition from weak to strong
angular momentum coupling.

The purely theoretical torque law for strongly-coupled stars in the MDM
is given in Eq. (5).
In the discussion so far, I have assumed that the mass-dependent part
of the torque is simply related to the star's convective turnover time $\tau$
by $f^2(B-V) = 2 \tau / k_I$ \citep{barneskim2010},
and also that the constant $K_{M1}$ is such as to give a 1-$M_{\sun}$
star the solar rotation period at the solar age.
There are thus only two free parameters in the pure form of the MDM:
the weak-coupling constant $K_{M0}$ and the transition timescale $\tau_M$
from the weakly- to the strongly-coupled modes.

The data do not place strong constraints on $K_{M0}$;
all that can be said is that it must be very small.
To estimate a limit I computed the value of $K_{M0}$ that degraded the fit for 
the 220-MYr cluster M34 so that the figure of merit $U_a$
decreased by 1.3 relative to $K_{M0}=0$ (corresponding to a 20-fold
decrease in the probability
that the observed and model period distributions were drawn from the same
parent distribution).
This procedure yields an upper limit of $K_{M0} \leq 2 \times 10^{30}$ g cm$^2$,
about 350 times smaller than $K_{M1}$.
Larger values of $K_{M0}$ degrade the fit with observations because they
are incompatible with the existence of the observed very fast rotators
(periods of order 0.1 day) in ZAMS-age clusters.
Applying the same criterion $\delta U_a = -1.3$ yields limits to the transition
timescale of 59 MYr $\leq \ \tau_M \ \leq$ 116 MYr, with the best-fit
$\tau_M$=80 MYr. 
The similarity between this timescale and the short timescale
($\tau/k_C)$ invoked by \citet{barnes2010} is very natural;
both arise from the need for $C$-sequence stars to disappear from
the $P_{rot}$-color diagrams for ages greater than a few hundred MYr.
On the other hand, the similarity between these timescales and the typical 
convection zone coupling times found for DZM models apparently is
coincidental.
The latter timescale is limited by different considerations, notably the need
to achieve near solid-body rotation by the age of the Sun, and
thereby agree with helioseismic data.

Using Barnes's \citep{barneskim2010} relation between the convective
turnover time $\tau$ and
$f^2(B-V)$, and the solar-rotation-consistent value for $K_{M1}$ 
(with individually optimized $K_{M0}$ and $\tau_M$) yields
a poor fit to the M34 data ($U_{M34} = -13.65$).
To learn whether plausible variations in the torque law might
give a significantly better fit, I generalized the strong-coupling
MDM torque law as follows:
\begin{equation}
{dJ \over dt} \ = \ S K_{M1} \left[ f^2(B-V) \right]^{1+\gamma} \Omega^3 \ \ ,
\end{equation}
where the exponent $\gamma$ permits varying the
mass-dependent shape of the torque function,
and $S$ is an arbitrary factor multiplying the coupling coefficient $K_{M1}$,
where $K_{M1}$ would yield the correct $P_{rot}$ at the solar age,
if $\gamma$ were zero.
To keep dimensions correct, I absorbed a factor of $[f^2(0.65)]^{-\gamma}$
(corresponding to $f^2$ evaluated at the solar color) into $S$.
Simultaneously adjusting the two parameters $S$ and $\gamma$ 
to maximize $U_{M34}$
yields $S = 1.2$, $\gamma = 1.0$, with the joint K-S probability
statistic $U_{M34} = -0.44$.
This model gives the solar
$P_{rot} = 26.7$ d at the solar age, slightly above the
sidereal Carrington period of 25.4 d.
This model is a good fit for all mass ranges;  
the K-S probabilities that
simulated data and the model are drawn from the same distribution
are \{0.83, 0.97, 0.60, 0.70\} for the 4 color bands.

Performing similar fits for five well-studied clusters of various ages gives  
the results in the top five lines of Table 1.
One can infer several systematic behaviors from this tabulation.
The fit quality as shown by the joint K-S probability metric $U_a$ is best for
M34 (age 220 MYr), and is worse for both younger and older clusters.
Indeed, for the youngest (Pleiades) and oldest (NGC 6811) clusters shown here 
the fits are poor,
with only one or two of the color bands reporting a K-S probability $U_{aj}$ 
better than
10\%, and the other bands being much worse.
Also, both fitted parameters as well as the inferred solar $P_{rot}$
at solar age show trends with cluster age.
These changes are in the sense that, for increasing age, the $I$-sequence
shape function of $(B-V)$
looks increasingly like a step function and less like a ramp, while
the projected solar rotation periods tend to decrease with increasing cluster
age.
Recall that, motivated by the SEM, the MDM assumes that 
$P_{rot}(B-V, t)$ is the product of $t^{-1/2}$ 
and an age-invariant function $f(B-V)$, which in turn is related to the
convective turnover time $\tau$.
Taken together, the fitting shortcomings just described suggest that  
none of these dependences are completely correct, though they happen to be
roughly so for clusters similar in age to M34.

\subsection{Comparison of DZM and SEM to MDM}

With the methodology just described, it is straightforward to
verify in a quantitative way that
the DZM and SEM provide relatively poor matches to observed cluster data.
(But note that the comparison statistics 
(e.g., $U_{M34}$) derive both
from model mis-fits to the $I$-sequence rotational
periods, and from poorly accounting for the $C$-sequence stars.
For current purposes the latter contribution is the more interesting one,
but, unfortunately, the former is usually larger.
The statistical results described in this section should thus be
considered indicative of differences among the models, but 
subject to misconstruction, nontheless.

To give a fair comparison with the MDM fits just described,
I generalized the DZM and SEM torque laws so that each of them includes
a scaling parameter $S$ and a $(B-V)$ shape exponent $\gamma$. 
I varied the values
of these parameters to give the best K-S probability, taken jointly over
the same color bands as used in the previous section.
The generalized torque laws I used were analogous to the one for the MDM.
For the DZM I tried two different torque laws, with different meanings
for the shape exponent $\gamma$.
First, for comparison with \citet{denissenkov2010},

\begin{eqnarray}
{dJ \over dt} \ = \ S K_W \ \Omega^3 \ \left [ {R \over M^{1+\gamma}} 
\right ]^{1/2},
\ \ \  \ \Omega \leq \Omega_{crit}, \\
\ \ \ \ \ \ \ \ \  = \ S K_W \ \Omega \Omega_{crit}^2 
\left [ {R \over M^{1+\gamma}} \right ]^{1/2},
\ \ \ \Omega \geq \Omega_{crit}, \nonumber
\end{eqnarray}

Second, for a more direct comparison with the radius-dependent torque law
of \citet{reiners2012}, I applied the $\gamma$ exponent to $R$, instead of $M$.
To agree with Eq. (5) of \citet{reiners2012}, I also changed the
explicit mass dependence to $M^{-2/3}$.
To distinguish this model from that of \citet{denissenkov2010}, I label it
DZM$_{\rm RM}$:

\begin{eqnarray}
{dJ \over dt} \ = \ S K_W \ \Omega^3 \ \left [ {R^{1+\gamma} \over M^{4/3}}
\right ]^{1/2},
\ \ \  \ \Omega \leq \Omega_{crit}, \\
\ \ \ \ \ \ \ \ \  = \ S K_W \ \Omega \Omega_{crit}^2
\left [ {R^{1+\gamma} \over M^{4/3}} \right ]^{1/2},
\ \ \ \Omega \geq \Omega_{crit}, \nonumber
\end{eqnarray}

Last, for the SEM:
\begin{equation}
{dJ \over dt} \ = \ - S I_* {\Omega^2 \over 2 \pi } \left [ {{2 \pi k_I} \over
 {\tau^{1 + \gamma} \Omega}} \ + \ {{\tau^{1+\gamma} \Omega} 
\over {2 \pi k_C}} \right ] ^{-1}.
\end{equation}

\begin{figure}[!hb]
\begin{center}
\epsscale{0.7}
\plotone{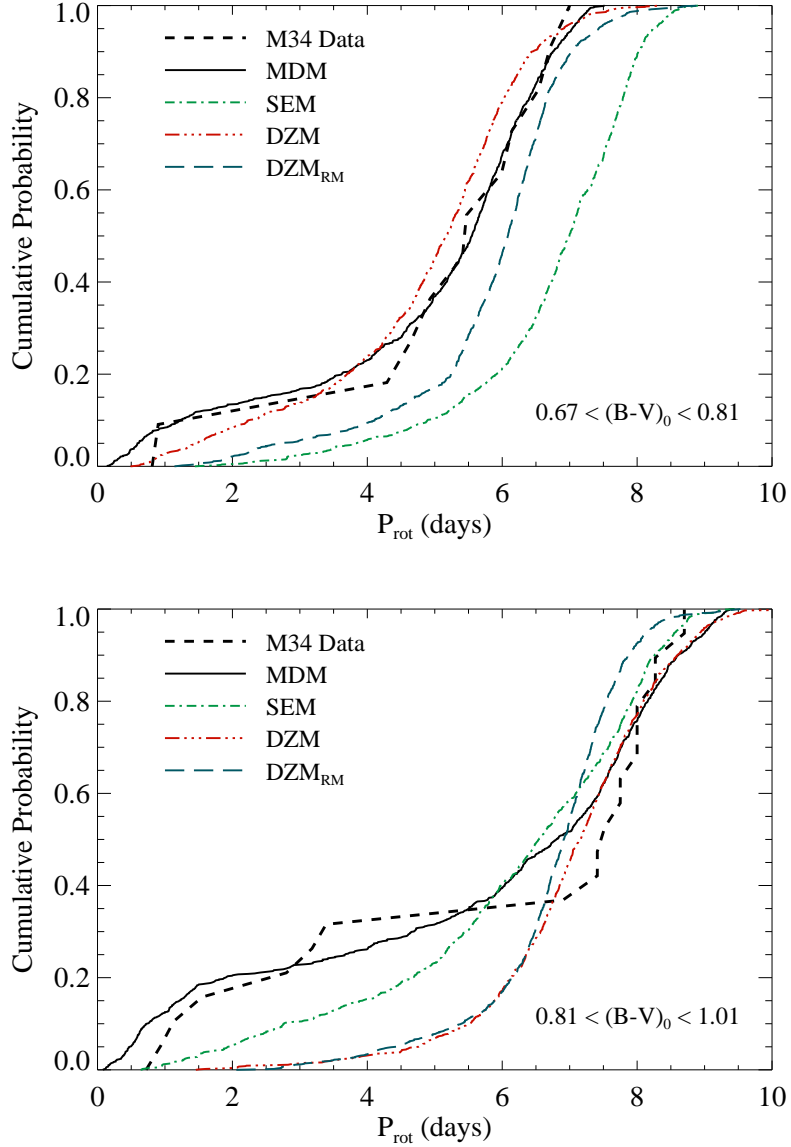}
\figcaption[fig8.eps]{Cumulative probability estimates shown as
functions of $P_{rot}$, for observations of the 220-MYr-old cluster
M34, 
and for population synthesis runs based on the DZM, DZM$_{\rm RM}$, 
SEM, and MDM.
The two panels show data summed over two different ranges of $(B-V)$ color,
as marked.
Observed periods are from \citet{meibom2009};  all modeling results are
for best-fit model parameters as described in the text.
\label{Figure 8}}
\end{center}
\end{figure}

The cluster M34 is a good choice for comparing model cumulative distributions, 
because it is
old enough that the $C$- and $I$-sequences are well
separated, but young enough that a significant fraction of $C$-sequence
stars remain.  
The DZM, DZM$_{\rm RM}$, and SEM fitting results for this cluster are shown 
in the last three lines of Table 1.
Figure 8 illustrates the cumulative probability distributions
on which the K-S statistics are based, for two of the color bands
used in the fits, for all (DZM, DZM$_{\rm RM}$, SEM, MDM) best-fit models.
Distributions for the other color bands are qualitatively similar.

For the DZM, the $U_{M34}$ statistic of -3.68 is fairly poor, though not 
as bad as for the worst clusters' (Pleiades, NGC 6811) fit with the MDM.
But the shape exponent's fitted value of 5.5 is far different from the
value suggested by theory ({\it i.e.,} roughly zero).
Moreover, the K-S probabilities for the individual color bands are
fairly good (about 0.45) only for the two bluemost colors, and are
much worse (about .025) for the two redmost colors. 

Allowing the torque to be an explicit function of radius (the DZM$_{\rm RM}$
model) gives a still poorer $U_{M34}$ statistic of -5.63.
The $\gamma$ value favored by the fit is 0.18. 
This corresponds to a torque law that scales as $R^{0.59}/M^{0.67}$,
which is quite different from the $R^{5.33}/M^{0.67}$ dependence derived by 
\citet{reiners2012}.

The $U_{M34}$ statistic for the SEM is considerably worse than for 
either version of the DZM,
and indeed is worse than for any cluster fitted using the MDM.
The SEM fit produces a K-S probability larger than 0.1 for only one
of the color bands; all the rest are smaller than 0.002.

For the DZM, DZM$_{\rm RM}$, and the SEM, these fitting statistics suggest that
one is attempting to force an inappropriate model onto the data, with
results that may be locally acceptable, but that fail in a global sense.
It is worth noting, however, that since most of the stars tabulated in
the M34 data lie on the $I$-sequence,
the fitting failures just described relate mostly to failures of the models
to reproduce the $(B-V)$ variation of the $I$-sequence.
Fig. 8 illustrates this;
particularly in its upper panel, the SEM and DZM cumulative probability curves 
reveal failure 
to match the $I$-sequence through displacements to right or left of the data
curve, for fairly long $P_{rot}$ and for cumulative probabilities larger 
than about 0.3.
But the shape of the $I$-sequence has only weak physical motivation
for all of the models. 
So it seems
likely that by choosing some other functional form for, say, the
$f(B-V)$ function that appears in the SEM, one could achieve
much better fits (although without necessarily adding substance 
to our physical understanding.)

At shorter periods and
lower cumulative probabilities, Fig. 8 shows something more essential.
In this part of the diagrams, the data curves have a step at periods
around 2 days, and an adjacent spread of larger periods 
over which the curve's slope is small.
These features are the signature of the $C$-sequence stars
at short periods and the paucity of stars at intermediate ones;
the MDM reproduces both features fairly well, but the other models do not.
This region of the $P_{rot}$ cumulative distributions offers the
strongest evidence in favor of the MDM, and at the same time suggests
ways in which the validity of the model might be tested in a quantitative way.

\section{Discussion}

\subsection{Summary of Conclusions}

In the foregoing sections, I have described a modeling exercise in which
I have tried to match observed $P_{rot}$-color diagrams using three
different models of the stellar spindown process.
In this attempt, I have assigned primary importance to the observed
existence in fairly young clusters of two sequences of stars, 
first noted by \citet{barnes2003}:
the $C$-sequence at $P_{rot}$ of about 2 days and less, 
separated by a poorly-populated gap from the $I$-sequence at longer periods.

The principal conclusion of the current modeling study is that previous
models (the double zone model, or DZM {\it e.g.} \citet{denissenkov2010}, and
Barnes symmetrical empirical model, or SEM \citep{barnes2010, barneskim2010}) 
do not adequately represent the best recent open cluster observations.
In particular, by virtue of the continuous way in which these models map
initial conditions into the history of torque on a star,
they appear unable to produce $C$-sequences that are as 
populous, well-defined, or at periods as short as those
seen in the observational data.
I propose the so-called Metastable Dynamo Model (MDM) as a solution to this
problem.
Its key hypothesis is that some or all stars go through an early phase in
which magnetic activity is present, but the angular momentum
coupling to the stellar wind is very small.
It appears to be quite difficult to explain the properties of
the $C$-sequence without invoking some such mechanism that decouples 
many young stars from the stellar wind torque, 
at least for a time.
Fits of the MDM parameters to cluster data constrain the strength of the
angular momentum coupling for the hypothetical weakly-coupled stars to be 
at least 300 times
less than for $I$-sequence stars.
The typical lifetime for the weakly-coupled phase is found to
be about 80 MYr for stars of 1 $M_{\sun}$;
the data are consistent with longer lifetimes of this phase for smaller-mass
stars, but the form of this dependence is poorly constrained. 
All three model classes (DZM, SEM, MDM) have failings, 
notably inability to reproduce
the time history of the color dependence of $P_{rot}$ on the $I$-sequence.
For the MDM, these inaccuracies are smallest for the cluster M34
(age about 220 MYr), and are larger for both younger and older clusters.

\subsection{Observational Tests}

The foregoing analysis suggests a number of observational tests that may
help choose among the various theories, test their premises, or refine their 
physical interpretations.

The DZM makes the striking prediction that most relatively
young stars (ages between 50 and 300 Myr, depending on mass) on the
$I$-sequence should have markedly different rotation rates in their CZs
and in their radiative interiors.
In principle, the radial variation of $\Omega$ can be measured with
asteroseismology.
This has been done for many years with
solar pulsations, and indeed,
using photometric data from the $Kepler$ mission,
recent analysis of pulsations in red giant stars has revealed large
differences in rotation rate between the tiny degenerate core and the
extensive convective envelope \citep{beck2012}. 
Asteroseismic measurements of rotation in young Sun-like stars have 
not yet been successful, however.
(\citet{deheuvels2014} have measured rotational splitting in $Kepler$
data for several subgiant stars.
But since the method used requires detection of mixed oscillation modes,
having properties of both pressure- and gravity-modes,
it is not applicable to the relatively young stars under discussion here.)
Young, magnetically active stars display larger photometric noise than
do their older, inactive brethren,
and moreover magnetic activity suppresses the surface amplitudes of 
acoustic oscillations, making them more difficult to 
observe \citep{chaplin2011}.
This unfortunate combination has so far prevented conclusive 
asteroseismic measurements of radial differential rotation in Sun-like stars.
Analysis of longer (full mission length) time series from $Kepler$ may
yet allow such measurements.

Apart from reproducing the morphology of the $C$- and $I$-sequences,
a successful model of rotational evolution should also give the observed 
distribution of stars in the gap between sequences.
For a uniform distribution of initial rotation periods, the MDM's prediction
in this respect is clear:
Stars in the gap must have recently transitioned from
weak- to strong-coupling modes,
hence they are rapidly evolving to longer periods;
as their periods increase, $dP_{rot}/dt$ decreases. 
Thus the star density
should be lowest at periods just above those in the $C$-sequence
and rise towards longer periods, up to the $I$-sequence.
This behavior is clearly visible in the probability density distributions
shown in Fig. 7.
Predictions for the DZM and the SEM are slightly more complicated,
and little work has been done concerning them,
but there are no fundamental obstacles to doing so.
The observational picture is more difficult, however.
Star counts between the $C$- and $I$-sequences 
are small, and as yet data 
are too sparse to make
strong tests of the models within the period gap.
Further observations are essential for this purpose.

Finally, if the MDM scenario is correct, then stars lying on 
the $C$-sequence must differ greatly from similar-mass stars in the period gap,
having much weaker angular momentum coupling to the stellar wind.
What could cause such a difference?
Clearly there must be some difference in the wind, 
in the magnetic field that threads it, or both.
Gross changes in the stellar wind seem implausible and in any case unobservable,
given current techniques.
If the difference is in the field, however, then diagnostics may be feasible.
Observations of X-ray luminosity as a function of $P_{rot}$
show no obvious step across the period range occupied by the $C$-sequence
\citep{pizzolato2003}.
Therefore the hypothetical discontinuous change in coupling
properties would likely be signalled not by a difference in the typical 
magnetic field strength, but rather by a change in its spatial organization.
This idea is not new;  it was suggested in \citet{barnes2003}, and has
since been explored using X-ray data by, e.g., \citet{wright2011}, 
by \citet{gondoin2012, gondoin2013}, 
with spectropolarimetry by, e.g., \citet{morin2011, jeffers2011}, and by
\citet{marsden2011a, marsden2011b}. 
and, in the context of ultra-cool dwarf stars, by combining X-ray and rotation
data \citep{cook2013}. 
A related line of argument starts with the observed Vaughan-Preston gap 
in the distribution of stars in $P_{rot}$-$R_{HK}^{\prime}$ space,
where $R_{HK}^{\prime}$ is the Mt. Wilson Ca II activity index
\citep{vaughan1980}. 
This gap was quickly interpreted as evidence for
a small discrete set of dynamo classes, most clearly seen
as relations between $P_{rot}$ and the dynamo cycle period
$P_{cyc}$ \citep{durney1981, brandenburg1998, bohm-vitense2007}.
In the current context, perhaps the most intriguing result of these studies
is the observation that some stars lie in a ``supersaturated dynamo'' state,
described by \citet{saar1999}.
Stars thus identified have short
rotation and cycle periods, and the power-law relation between these periods
has the opposite sign from that seen in stars with longer $P_{rot}$.
Moreover, the transition between the supersaturated and other dynamo modes
appears to be abrupt (because there are very few transitional objects seen), 
and involves a discontinuous change in the cycle period.

The various lines of study just described involve a wide range of
stellar circumstances and several different (and hypothetical)
formulations of dynamo physics.
So it is not clear that all of these studies relate to the same phenomena,
or are governed by the same processes.
Nevertheless, taken together they reinforce the idea that 
a variety of dynamo modes 
might exist, yielding (among other properties) different partitioning
of power across large and small spatial scales.
If almost all of the magnetic energy were found in small-scale strucures
in which positive and negative field regions accurately cancel one another,
then little field might penetrate to heights where the stellar wind expansion
begins.
The result would be an almost field-free wind, and only minimal torque on
the star.
Differences among photospheric spatial structures
may be identifiable using Doppler imaging and spectropolarimetry,
by comparing very rapid rotators against stars with similar 
activity diagnostics but slower rotation periods.
Measuring the cycle periods of very fast rotators might also be revealing,
as a probe of deeper-seated differences among the properties of stellar dynamos.

\subsection{Final Considerations}

It is worthwhile to reiterate a few points, and to raise some
issues for further work.
(1) The population synthesis calculations described above 
suggest that the MDM has some validity,
but they are by no means conclusive evidence that it is correct,
nor that the DZM or SEM are wrong.
There may well be parameter choices for these latter models that will
better reproduce the $C$-sequence population than the ones I have
employed here.
Better fitting of the $I$-sequence is almost surely feasible
for all model types.
(2) Within the MDM framework, it appears that models 
enforcing solid-body rotation
match the young-cluster observations,
but there is as yet nothing to show that differentially-rotating models
are excluded.
(3) The basic MDM assumes that all stars begin life in the weakly-coupled
dynamo mode.
But in very young clusters there is evidence that the MDM's assumptions
place too many stars on the $C$-sequence.
Thus, perhaps only a fraction (perhaps as few as half) of all stars
initially occupy the weak-coupling  mode. 
A statistical test of this conjecture may be feasible with the data
that are presently in hand.
(4) All model classes have difficulty fitting the shapes of $I$-sequences
with age, especially for stars older than a few hundred MYr.
Indeed, given this difficulty,
the $\Omega^3$ torque law (hence the $t^{-1/2}$ period law) 
may be only an approximation.
For recent evidence that this is so, based on stars with measured $P_{rot}$
and asteroseismic ages, see \citet{metcalfe2014}.
Better modeling may require a function of both $\Omega$ and mass;
making a useful guess about the form of such a function will be 
difficult, lacking a physical picture of the important processes.

I am grateful to Sydney Barnes and to Marc Pinsonneault for conversations
that inspired and informed this work, and to Travis Metcalfe, David Soderblom,
Soeren Meibom, Lynne Hillenbrand, and to the anonymous referees for their 
useful comments on 
early versions of this paper.

\clearpage


\include{table1}

\end{document}

%% file: table1.tex
\begin{table}[!hb]
\begin{center}

\caption{Results of K-S Fits to Cluster $P_{rot}$ Data}

\begin{tabular}{lcrcccc}
\tableline\tableline
Cluster & Age & Nstars & Model & Scale & Exponent & FOM \\
 & (MYr) &  & Class & $S$ & $\gamma$ & $U_a$ \\
\tableline
Pleiades & 125 & 147 & MDM & 1.37 & 1.50 & -5.15 \\
M35 & 150 & 172 & MDM & 1.36 & 0.75 & -2.22 \\
M34 & 220 & 70 & MDM & 1.20 & 1.00 & -0.44 \\
M37 & 550 & 187 & MDM & 0.84 & 0.45 & -3.60 \\
NGC 6811 & 1000 & 48 & MDM & 0.88 & -0.46 & -7.52 \\
M34 & 220 & 70 & DZM & 1.21 & 5.50 & -3.68 \\
M34 & 220 & 70 & DZM$_{\rm RM}$ & 1.99 & 0.18 & -5.63 \\
M34 & 220 & 70 & SEM & 1.60 & 0.10 & -9.54 \\
\tableline
\end{tabular}
\tablecomments{The sources for $P_{rot}$ data for the various clusters
are  (Pleiades: \citet{hartman2010}), (M35: \citet{meibom2009}),
(M34: \citet{meibom2011a}), (M37: \citet{hartman2009}), (NGC 6811:
\citet{meibom2011b}).
}
\end{center}
\end{table}